\documentclass[sigconf]{acmart}

\AtBeginDocument{%
  }

\usepackage{physics}
\usepackage{tikz}
\usepackage{quantikz}

\copyrightyear{2026}
\acmYear{2026}
\setcopyright{cc}
\setcctype{by}
\acmConference[SCA/HPCAsia 2026]{SCA/HPCAsia 2026: Supercomputing Asia and International Conference on High Performance Computing in Asia Pacific Region}{January 26--29, 2026}{Osaka, Japan}
\acmBooktitle{SCA/HPCAsia 2026: Supercomputing Asia and International Conference on High Performance Computing in Asia Pacific Region (SCA/HPCAsia 2026), January 26--29, 2026, Osaka, Japan}
\acmPrice{}
\acmDOI{10.1145/3773656.3773689}
\acmISBN{979-8-4007-2067-3/2026/01}

\begin{document}

\title{GCAMPS: A Scalable Classical Simulator for Qudit Systems}

\author{Ben Harper}
\authornote{Both authors contributed equally to this research.}
\email{harperb@unimelb.edu.au}
\affiliation{%
  \institution{School of Physics, The University of Melbourne}
  \city{Parkville}
  \state{Victoria}
  \country{Australia}
}
\affiliation{%
  \institution{Data61, CSIRO}
  \city{Clayton}
  \state{Victoria}
  \country{Australia}
}

\author{Azar C. Nakhl}
\authornotemark[1]
\email{chris.nakhl@unimelb.edu.au}
\affiliation{%
  \institution{School of Physics, The University of Melbourne}
  \city{Parkville}
  \state{Victoria}
  \country{Australia}
}

\author{Thomas Quella}
\affiliation{%
  \institution{School of Mathematics \& Statistics, The University of Melbourne}
  \city{Parkville}
  \state{Victoria}
  \country{Australia}
}

\author{Martin Sevior}
\affiliation{%
  \institution{School of Physics, The University of Melbourne}
  \city{Parkville}
  \state{Victoria}
  \country{Australia}
}

\author{Muhammad Usman}
\affiliation{%
  \institution{School of Physics, The University of Melbourne}
  \city{Parkville}
  \state{Victoria}
  \country{Australia}
}
\affiliation{%
  \institution{Data61, CSIRO}
  \city{Clayton}
  \state{Victoria}
  \country{Australia}
}

\renewcommand{\shortauthors}{Harper et al.}

\begin{abstract}
    Classical simulations of quantum systems are notoriously difficult computational problems, with conventional state vector and tensor network methods restricted to quantum systems that feature only a small number of qudits. The recently introduced Clifford Augmented Matrix Product State (CAMPS) method offer scalability and efficiency by combining both tensor network and stabilizer simulation techniques and leveraging their complementary advantages. This hybrid simulation method has indeed demonstrated significant improvements in simulation performance for qubit circuits. Our work generalises the CAMPS method to higher quantum degrees of freedom --- qudit simulation, resulting in a generalised CAMPS (GCAMPS). Benchmarking this extended simulator on quantum systems with three degrees of freedom, i.e. qutrits, we show that similar to the case of qubits, qutrit systems also benefit from a comparable speedup using these techniques. Indeed, we see a greater improvement with qutrit simulation compared to qubit simulation on the same $T$-doped random Clifford benchmarking circuit as a result of the increased difficulty of conventional qutrit simulation using tensor networks. This extension allows for the classical simulation of problems that were previously intractable without access to a quantum device and will open new avenues to study complex many-body physics and to develop efficient methods for quantum information processing.
\end{abstract}

\ccsdesc[500]{Computing methodologies~Quantum mechanic simulation}
\ccsdesc[300]{Computing methodologies~Simulation evaluation}
\ccsdesc[500]{Computing methodologies~Simulation tools}

\keywords{Tensor Networks, Matrix Product States, Stabilizer Simulation, Qudit Simulation}

\maketitle

\section{Introduction}

Classical simulation of quantum systems continues to underpin research in quantum information and computation~\cite{nakhl_stabilizer_2025,dang_distributed_2017,aaronson_improved_2004_2,goh_lie-algebraic_2025,feng_quon_2025} as well as in the study of quantum many-body physics~\cite{orus_tensor_2019,werner_positive_2016_2} and topological materials~\cite{mortier_tensor_2022, ran_tensor_2020}, among others. However, it is a notoriously hard computational problem that scales poorly with respect to the quantum system size (i.e. the number of qubits or qudits). Therefore, the development of scalable and efficient classical simulation frameworks for quantum systems and their performance benchmarking in terms of quantum system size or complexity is an active area of research~\cite{nakhl_stabilizer_2025,pashayan_fast_2022,lami_quantum_2024,qian_augmenting_2024,fux_disentangling_2025}.   

Currently classical simulators can be classified into a number of different categories depending on the class of quantum systems that they cannot efficiently simulate. Whereas paradigmatic simulators based on the simulation of state vectors or density matrices scale poorly with the number of sites~\cite{nielsen_quantum_2012}, there exist several sophisticated techniques that scale with other quantum resources. The most common among these are tensor networks~\cite{schollwock_density-matrix_2011,orus_practical_2014} that scale with entanglement~\cite{nakhl_calibrating_2024,xu_herculean_2023}, and stabilizer tableaus~\cite{gottesman_stabilizer_1997,aaronson_improved_2004_2} that scale with magic~\cite{bravyi_improved_2016,bravyi_simulation_2019}. Recently, a concerted effort has been made in unifying these two simulation methods, stabilizers and tensor network methods, with the aim of harnessing the benefits of both~\cite{masot-llima_stabilizer_2024,lami_quantum_2024,qian_augmenting_2024,nakhl_stabilizer_2025,fux_disentangling_2025}. To that end, it has been demonstrated that such methods are capable of simulating systems which have an extensive amount of both entanglement and magic~\cite{fux_disentangling_2025,nakhl_stabilizer_2025,liu_classical_2025} and have been used in the simulation of quantum many-body systems and their evolution~\cite{qian_clifford_2025,mello_clifford_2025,frau_stabilizer_2024,fan_disentangling_2025}. 

In order to build an extensive quantum many-body simulator, however, one must support quantum mechanical degrees of freedom of dimension $>2$ (noting that $d=2$ refers to qubit systems), which generally for conventional simulators is not a significant undertaking. Indeed, for a conventional stabilizer simulator such an extension is typically not considered as simulations of interest enabled by such an extension will in general require universality, with universal \textit{extended} stabilizer simulators scaling strictly exponentially with the number of non-Clifford operations performed, that is regardless of the actual non-stabilizerness of the final state~\cite{bravyi_simulation_2019}. Stabilizer Tensor Network (STN) methods however, like conventional tensor networks, are strictly universal and have a resource cost that need not scale exponentially with any given quantum operation.

In this work, we introduce the \textit{Generalised Clifford Augmented Matrix Product State (GCAMPS)} classical simulation method, a generalisation to the recently reported Clifford Augmented Matrix Product State (CAMPS) method~\cite{lami_quantum_2024,qian_augmenting_2024} to support arbitrary quantum degrees of freedom. We provide a detailed outline on the qudit Clifford group and how stabilizer tableau simulation could be extended to support the simulation of operators within this group. We then describe the GCAMPS method and show how one may in principle simply slot in a qudit stabilizer simulator to have a universal quantum simulator. We outline the process of finding appropriate disentanglers, a set of operators required for efficient GCAMPS simulation. We also note that how in general finding and utilising these operators does not scale efficiently, restricting the utility of generalised GCAMPS to low quantum degrees of freedom. Finally, we demonstrate the efficacy of this extension by simulating $T$-doped random Clifford circuits for qutrits ($d=3$), demonstrating similar scaling behaviour to known results for qubits~\cite{nakhl_stabilizer_2025,fux_disentangling_2025,liu_classical_2025}. Our results show that the GCAMPS framework demonstrates a significant improvement in the memory usage and runtime compared to conventional MPS simulation in the regime where the $T$ gate count is less than the number of qudits. Notably, this improvement over MPS for qutrit simulations is greater than that seen with the qubit simulations.  

Our work is one of the first concerted efforts to develop a scalable and efficient classical simulator for qudit quantum systems, which could enable future investigations of complex many-body physics~\cite{lauchli_spin_2006} and magic in quantum systems with higher-dimensional state space beyond qubits~\cite{zhang_quantum_2024,magni_quantum_2025,wang_stabilizer_2023,turkeshi_magic_2025}, uncovering new avenues for quantum information processing.    

\section{Preliminaries}
In this section, we will provide a brief overview of the relevant concepts and definitions. We will start with basic definition of qubits and qudits in the state vector representation. We will then provide an overview of stabiliser simulation techniques for both qubits and qudits, followed by a brief overview tensor networks and Matrix Product States (MPS). Finally, we will describe GCAMPS and introduce how this may be generalised to arbitrary quantum degrees of freedom.

\subsection{Defining Qubits and Qudits}
Pure quantum systems are described by a state vector, where each element is a complex amplitude associated with a particular basis state~\cite{nielsen_quantum_2012}. For a 2-level system (a qubit), the state vector $\ket{\psi}$ is given by,

\begin{equation}
    \ket{\psi} = \alpha \ket{0} + \beta \ket{1} =
    \begin{bmatrix}
        \alpha \\
        \beta
    \end{bmatrix}.
\end{equation}
where $\alpha, \beta \in \mathbb{C}$ and $|\alpha|^2 + |\beta|^2 = 1$ to ensure normalisation.

\begin{figure}
    \includegraphics{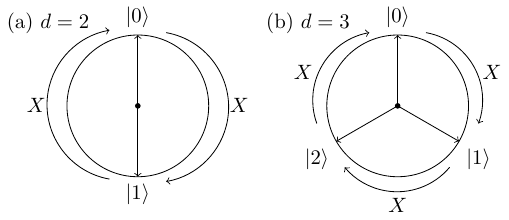}
    \caption{The qudit stabilizer states correspond to positions on a $d$-digit clock, where the Pauli $X$ operator moves the clock's hand. For (a) $d=2$ this corresponds to up and down spins. (b) When $d \geq 3$ there are more possible states. Similarly, $Z$ and $Y$ rotate between $d$ orthogonal basis states.}
    \label{fig:clock}
\end{figure}

This representation generalizes to qudits ($d$-level systems) where the state vector has $d$ complex amplitudes corresponding to the $d$ orthonormal basis states. These states may be represented as points on a clockface, as in Fig~\ref{fig:clock}.
\begin{equation}
    \ket{\psi_d} = c_0 \ket{0} + c_1\ket{1} + \ldots + c_{d-1} \ket{d-1} = \begin{bmatrix}
        c_0 \\
        c_1 \\
        \vdots \\
        c_{d-1}
    \end{bmatrix}
\end{equation}
Where qubits are the quantum analog to classical binary logic, qudits are the quantum analog to ternary or higher dimensional logic. A system of multiple qudits is formed via the tensor product of the subsystems, e.g. $\ket{\psi_{A+B}} = \ket{A} \otimes \ket{B}$ and linear combinations of such states.

Quantum operations correspond to unitary matrices acting on these state vectors. As such, linear algebra provides the mathematical framework for simulating quantum systems. However, direct matrix multiplication scales exponentially with the size of the system --- the state vector of an $n$-qudit system has size $\mathcal{O}(d^n)$, and the corresponding unitary operations are of size $\mathcal{O}(d^{2n})$. This makes direct state vector simulation infeasible for systems beyond a modest number of qudits. As a result, practical simulation of quantum systems relies on more efficient representations of quantum states and the operations that act upon them.

\subsection{Stabilizer Simulation for Qubits ($d=2$)}
One such representation is the stabilizer formalism~\cite{gottesman_stabilizer_1997,aaronson_improved_2004_2}. The Pauli matrices $X$, $Y$ and $Z$ (defined in equation~\ref{paulis2}), along with the identity matrix $I$, generate the Pauli group on a single qubit.

\begin{equation}
    X = \begin{bmatrix}
        0 & 1 \\
        1 & 0
    \end{bmatrix} \qquad
    Z = \begin{bmatrix}
        1 & 0 \\
        0 & -1
    \end{bmatrix} \qquad
    Y = iXZ = \begin{bmatrix}
        0 & -i \\
        i & 0
    \end{bmatrix}\label{paulis2}
\end{equation}

\noindent
By taking tensor products of these matrices across multiple qubits, we form Pauli strings. For instance, the application of a Pauli-$X$ on qubit 2 and Pauli-$Z$ on qubit 4 with no action on the other qubits is represented by the Pauli string $P=IXIZI$ for a 5-qubit system. Given a quantum state $\ket{\phi}$ it is possible to find the Pauli strings $S_i$ corresponding to operators which \textit{stabilize} the state; that is, operations which act like the identity on the state $\ket{\phi}$.
\[ S_i \ket{\phi} = \ket{\phi} \]
The group of states formed by the stabilizers of a state form a subgroup of the Pauli group, known as the stabilizer group $\mathcal{S}$.

If an $n$ qubit state possess a set of $n$ independent, commuting stabilizers, that state is the unique quantum state (up to a physically insignificant global phase) stabilized by its stabilizer group. This state can therefore be represented by storing the $n$ stabilizers in a stabilizer tableau, which requires only $\mathcal{O}(n^2)$ memory --- significantly more efficient than storing a full state vector. The tableau is typically represented as a binary matrix where each row corresponds to a Pauli string (stabilizer or destabilizer). For example, the state $\ket{00}$ is stabilized by the strings $ZI$ and $IZ$, and its tableau $M$ is,
\begin{equation}
     M = \left[\begin{array}{cc|cc|c}
        Z & I & I & I & +1 \\
        I & Z & I & I & +1 \\
        \hline
        I & I & X & I & +1 \\
        I & I & I & X & +1
    \end{array}\right]
\end{equation}
In practice, the $X$ and $Z$ components are stored separately as binary values, as in the tableau above. The top half of the tableau encodes the stabilizer generators, while the bottom half encodes the destabilizer generators --- operators which anticommute with one or more stabilizers. Together these $2n$ operators form a complete basis for the Pauli group on $n$ qubits. An additional column is used to store the global phase of each generator ($\pm 1$ for qubits), resulting in a tableau of size $2n \times 2n+1$.

To simulate quantum operations, we must update the stabilizer tableau under unitary transformations.  Given a quantum state $\ket{\phi}$ and a unitary gate $U$, we wish to find the tableau corresponding to the state $U\ket{\phi}$. Since the stabilizers satisfy $S\ket{\phi} = \ket{\phi}$, we have,
\begin{align*}
    U\ket{\phi} &= U S \ket{\phi}, \qquad S \in \mathcal{S} \\
    &= U S U^\dag U \ket{\phi} \\
    &= S' (U\ket{\phi})
\end{align*}
where $S' = U S U^\dag$ is the transformed stabilizer. Thus, simulating an operation corresonds to updating each stabilizer generator by conjugating it with the gate $U$.

The set of gates that map Pauli strings to other Pauli strings under conjugation are known as the Clifford gates. These include the Pauli gates $X, Y, Z$, Hadamard, phase gate $\sqrt{Z}$, and the entangling CNOT gate. Gates outside the Clifford group, known as non-Clifford gates, do not preserve the Pauli group and cannot be directly simulated within the stabilizer formalism. Although the Clifford gates do not form a universal gate set, many important quantum circuits (such as quantum error correction (QEC) circuits) are made up entirely of Clifford gates. Hence, stabilizer simulation is essential in studying these systems. The classical simulability of Clifford circuits is known as the Gottesman-Knill theorem~\cite{gottesman_stabilizer_1997}.

\subsection{Generalisation to Qudits ($d>2$)} \label{sec:stab}
While stabilizer simulation for qubits ($d=2$) is well studied, its generalisation to higher dimensions has received limited attention. Despite this, the stabilizer formalism naturally extends to higher dimensions~\cite{brandl_efficient_2024,Nguyen_Sdim_A_Qudit_2025,Wang_2020}. The generalised Pauli operators for a single qudit of dimension $d$ are defined by their action on computational basis states as,
\begin{align*}
    \omega = \exp\left( \frac{2\pi i}{d}\right) &\qquad XZ = \omega^{-1} ZX \\
    X\ket{j} = \ket{j+1 \mod d}, \qquad
    &Z\ket{j} = \omega^j \ket{j}, \qquad
    Y = XZ.
\end{align*}
Note that for even values of $d$, the $Y$ gate requires an extra phase factor $\tau = \omega^{1/2}$ such that $Y = \tau XZ$. This guarantees that each of the Pauli operators is of order $d$, i.e. $P^d = I$ for $P \in\{X, Y, Z\}$. We also note that all qudit operators may be written as a linear combination of $X^x Z^z$ terms, as these form a basis for the space of unitary matrices.

The generalised stabilizer tableau retains the same form as in the qubit ($d=2$) case, but its entries must now be integers over $\mathbb{Z}_d$. Each element of the tableau now corresponds to the exponent of a Pauli operator, such as $X^a Z^b$, where $a, b \in \mathbb{Z}_d$ denote an exponent of each Pauli, and the overall phase of each stabilizer row is now an element of the multiplicative group generated by $\omega$, i.e. $\omega^k$ for $k \in \mathbb{Z}_d$. Consequently, the tableau is still a $2n \times 2n + 1$ matrix, but its entries are integers modulo $d$, and the phase column tracks elements of $\mathbb{Z}_d$ instead of just $\pm 1$.

To perform stabilizer simulation, we must update the tableau under Clifford operations --- unitaries that map Pauli operators to other Pauli operators under conjugation. The generalized Clifford gates include the single qudit $H$ and $S$ gates,

\[ H_d = \frac{1}{\sqrt{d}} \sum_{i=0}^{d-1} \sum_{j=0}^{d-1} \omega^{ij} \ket{j}\bra{i}\;, \]
\[ S_{d_\text{odd}} = \sum_{j=0}^{d-1} \omega^{\frac{j(j-1)}{2}} \ket{j}\bra{j}\;, \qquad
S_{d_\text{even}} = \sum_{j=0}^{d-1} \tau^{j^2} \ket{j}\bra{j}, \]
and the two-qudit entangling gate,
\[ \text{SUM}_d \ket{i, j} = \ket{i, (i + j) \mod d}. \]
which is the qudit generalisation of the CNOT gate. These gates act on Pauli operators via conjugation, and the update rules for the tableau are defined in table~\ref{tab:update_rules}.

\begin{table}
    \centering
    \begin{tabular}{cc}
         \toprule
         $H_d$ & $X \to Z$ \\
             & $Z \to X^{-1}$ \\
         \midrule
         $S_d$ & $X \to (\tau) XZ = Y$ \\
             & $Z \to Z$ \\
         \midrule
         $CNOT_d$ & $X \otimes I \to X \otimes X$ \\
             & $I \otimes X \to I \otimes X$ \\
             & $Z \otimes I \to Z \otimes I$ \\
             & $I \otimes Z \to Z^{-1} \otimes Z$
    \end{tabular}\caption{Tableau update rules for qudit stabilizer simulation}\label{tab:update_rules}
\end{table}

\subsubsection{Decomposing Operators}  \label{sec:operators}
In the hybrid simulation scheme described below, it is necessary to express arbitrary Pauli operators in terms of the (de)stabilizer generators. That is, given a Pauli string represented as a product of $X$ and $Z$ on each qubit (where $Y$ = $XZ$), in a system of $n$ qudits,
\[ P = \prod_{i = 1}^{n} X_i^{x_i} Z_i^{z_i} \]
where $x_i, z_j \in \mathbb{Z}_d$, we aim to express $P$ as a product of the (de)stabilizer generators $\{S_j\} \bigcup \{D_{j}\}$ raised to integer powers,
\[ P = c\prod_j S_j^{s_j} D_j^{d_j} \]
where each $s_j, d_j$ is an exponent of the (de)stabilizer generator $S_j, D_j$, and $c$ is an overall phase. To compute the coefficients $s_j, d_j$, we perform Gaussian elimination over the field $Z_d$ to solve the following system of linear equations,
\begin{equation}
    P = M \cdot \left[\begin{array}{c} s \\ \hline d \end{array}\right]\;,
\end{equation}
where $M$ is the $2n \times 2n$ stabilizer tableau with the phase column excluded, and $P$ is the Pauli string we wish to find a decomposition for. The solution vector contains the required exponents of the tableau rows to produce the string $P$ (up to a global phase). To determine the overall phase factor $c$ of the decomposition, we explicitly multiply the tableau rows according to the coefficient determined above, while keeping track of the cumulative phase resulting from the non-commutation multiplication of Pauli operators.

\subsubsection{Universal Quantum Simulation}
To represent arbitrary unitary operations and achieve universal quantum computation, the Clifford gates must be extended with the addition of a single non-Clifford gate. A common choice is the $T = \sqrt{S}$ gate, also called the $\pi/8$ gate. While a convenient general form of the $T$ gate does not exist, for the cases of $d=2, 3$ considered in simulation in this work, the $T$ gate takes the form,
\[
    T_2 = \begin{bmatrix}
        1 & 0 \\
        0 & e^{i\pi / 4}
    \end{bmatrix}, \qquad
    T_3 = \begin{bmatrix}
        1 & 0 & 0 \\
        0 & e^{i\pi/9} & 0 \\
        0 & 0 & e^{8i\pi / 9}
        
    \end{bmatrix}
\]
Any unitary quantum gate may be represented by a decomposition of gates from the set $\text{Cliffords}\ \bigcup\ \{ T \}$~\cite{nielsen_quantum_2012}. We note that the T gate could be replaced by an arbitrary non-Clifford gate to achieve universal quantum computation~\cite{nebe2006self}. However, the $T$ gate is typically chosen as it is most suitable in fault-tolerant quantum computing protocols. For the purposes of classical simulation protocols it is typical to instead pick a parametrised rotation gate as the non-Clifford gate as this generally leads to simpler circuits. Indeed, our implementation of the simulation method introduced below utilises the parametrised $R_Z(\theta)$ gate~\cite{nielsen_quantum_2012} where $T=e^{i\pi/8}R_Z(\pi/4)$. In general, the choice of non-Clifford does not greatly affect the simulation cost of the (G)CAMPS simulation method introduced below as an optimisation routine is performed after the application of such gates.

\subsection{Tensor Network Simulation}

\begin{figure}
    \centering
    \includegraphics[width=0.9\linewidth]{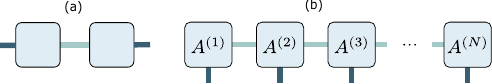}
    \caption{(a) A tensor network representation of the product $AB$, i.e.\ matrix multiplication. (b) The graphical representation of the MPS in Equation~\eqref{eq:mps}}.
    \label{fig:tn-fig}
\end{figure}

Tensor Networks~\cite{orus_practical_2014} are a decomposition of many-body quantum systems across multiple tensors (n-dimensional arrays). Graphically they may be represented using graphs in which each node represents a tensor and edges represent a sum over all elements along a given axis between the two tensors, see Figure~\ref{fig:tn-fig}(a). Performing this summation results in the two tensors becoming one (an operation known as a contraction). Dangling edges in this graphical representation correspond to the physical indices of the system, i.e. after the matrix multiplication in Figure~\ref{fig:tn-fig}(a) is performed, the outermost dimensions represented by the dangling edges on the left and right correspond to the resulting matrix shape.

Matrix Product States (MPS)~\cite{schollwock_density-matrix_2011,orus_practical_2014} are a one-dimensional tensor network, i.e. graphically represented as a path graph with an extra dangling edge at each node, see Figure~\ref{fig:tn-fig}(b). They are particularly noteworthy as contraction, which is required for evolution and evaluation of the system, may be performed efficiently. Mathematically, we may write down an MPS for a many-body state each with quantum degrees of freedom $d$ as,
\begin{equation}
    \ket{\psi} = \sum_{s_1,s_2,\ldots,s_N} A_{s_1}^{(1)} A_{s_2}^{(2)} \dots A_{s_N}^{(N)}\ket{s_1s_2\dots s_N} \label{eq:mps}
\end{equation}
where the $s$ are a $d-$nary string of some number $0,\dots,d^N -1$ with the $\ket{s}$ representing each of the basis states of the system. The $A_{s_i}^{(i)}$ are of some size $\chi_{i-1} \times \chi_{i}$,
with the matrices at the end being of size $1 \times \chi_1$ and $\chi_{N} \times 1$ respectively. $\chi_i$ is referred to as the bond-dimension (between sites $i$ and $i+1$) and crucially describes the bipartite entanglement of the state at that site.
One may find a lower-entanglement approximation to the state $\ket{\psi}$ by performing a Singular Value Decomposition (SVD) on the $A_{s_i}^{(i)}$, keeping only a fixed number of singular values.

Evolving an MPS, or in fact any tensor network involves the contraction of an operator, that is a tensor or tensor network with $2n$ physical indices along $n$ indices of the original tensor network. Operators that act on only one tensor in the network do not increase any of the bond-dimensions of the tensor network and hence the memory cost of the operations is unchanged. Operators that do act on multiple tensors however may increase the bond-dimension of the relevant edges by a factor of $d$. Hence, it is preferable to minimise the number of multi-site operations and ensuring any multi-site operations act on adjacent sites as to minimise the region in which the bond-dimension is possibly increased.  

\section{GCAMPS}
\begin{figure*}
    \includegraphics{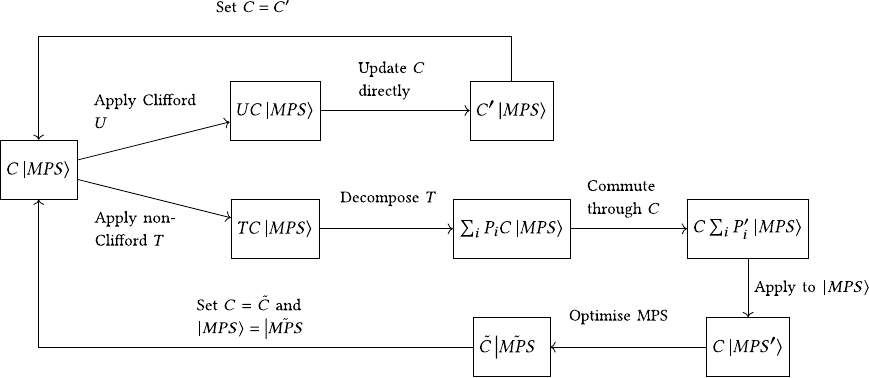}
    \caption{GCAMPS Workflow: The state is represented by an MPS $\ket{\text{MPS}}$ and a leading Clifford operation $C$ represented by a stabilizer tableau. (Top loop) Clifford operations directly update the tableau resulting in the top loop. (Bottom loop) Non-Clifford operations must first be decomposed into a sum of Paulis $\sum_i P_i$ and commuted through the tableau. The commuted operator is the applied directly to the MPS, after which the MPS may be optimised by extracting entanglement using Clifford operations that are applied to the stabilizer tableau. GCAMPS extends CAMPS by incorporating a qudit stabilizer simulator for the Clifford $C$ and extending the decomposition and optimisation steps in the bottom loop to utilise the qudit Cliffords.}\label{fig:1}
\end{figure*}

One may hybridise the two simulation methods introduced earlier by augmenting the MPS by a Clifford that is efficiently representable using the stabilizer tableau formalism introduced in Section~\ref{sec:stab}. For qubits, this hybrid simulation method is referred to as the Clifford Augmented Matrix Product State (CAMPS) method~\cite{lami_quantum_2024,qian_augmenting_2024}. Here, we specifically outline how a generalised CAMPS (GCAMPS) works for arbitrary quantum degrees of freedom. Figure~\ref{fig:1} illustrates a flowchart representation of the GCAMPS simulation framework.

States represented in the (G)CAMPS formalism are of the form $\ket{\psi} = C\ket{\text{MPS}}$, where $C$ represents a Clifford operation that ideally ``contains'' all the entanglement in the state, leaving the MPS in a state close to a product state. For quantum states represented in this way, performing a Clifford gate simply updates the Clifford $C$ via a tableau update.  To perform non-Clifford operations one needs to find the decomposition of said operator of the form $U=\sum_j c_{j} \prod_{i=1}^n X_i^{x_j(i)} Z_i^{z_j(i)}$ which may be found numerically by solving a linear system, given $U$ acting over a small number of sites. The procedure in Section~\ref{sec:operators} then finds how this decomposed operator $U$ commutes through $C$. After commutation, the commuted operator $\tilde{U}$ may be applied directly to the MPS. After commutation, the commuted operator $\tilde{U}$ may be applied directly to the MPS. While the original operator may have been local, the commutated operator in general will not be, resulting in a non-local operation on the MPS which may increase the bond-dimension across certain sites.

Having performed a non-Clifford operation, one may seek to reduce the entanglement in the MPS which is likely to increase after performing a non-Clifford gate on an entangled system. This may be done by heuristically applying a Clifford $Q$ to the MPS that reduces the entanglement in the system, one may then preserve the state by updating $C$ as $\tilde{C}= CQ^\dagger$. For a simple search where one seeks to apply only two-qudit Clifford operations there are only $20$ uniquely entangling Cliffords for $d=2$ and $90$ uniquely entangling Cliffords for $d=3$ so in these instances this search may be done exhaustively for any single layer made up of two qudit operators.

An alternative to the entanglement reduction procedure described here is to instead prepare a resource state which generally have a fixed computational resource cost to prepare, and inject that into the data-register of the circuit. The resulting simulation methods is referred to \textit{Magic State Injected Stabliizer Tensor Networks} (MAST) in the literature~\cite{nakhl_stabilizer_2025}. In the MAST setup, the computational resource cost results from the projection step of the injection gadget (see Figure~1 of~\cite{nakhl_stabilizer_2025}). For gate based quantum simulations where one represents their state largely in terms Clifford operations and local non-Clifford operations this approach may be preferred over CAMPS based methods which require the above described optimisation routine which may be computationally inefficient.

The computation of Pauli observables in $d=2$ in this framework is rather simple as Pauli strings transform to other Pauli strings when passing through $C$, resulting in an expectation value of the form $\bra{\text{MPS}}\prod_iX_i\prod_jZ_j\ket{\text{MPS}}$ which may be efficiently computed with MPS. For $d\neq 2$ this task becomes more difficult however as the generalised Pauli operators are not Hermitian and hence not a quantum observables. Typically, then one defines the observable $O_\sigma = \frac{\sigma + \sigma^\dagger}{2}$, which is related to the generalised Paulis and is Hermitian. Computing the observable $O_\sigma$ for $d\neq 2$ may at worst increase the bond dimension $\chi$ by a factor of $d$.

\subsection{Computing the Disentanglers}
The entanglement-reduction step described above requires identifying the uniquely entangling Cliffords, allowing for an exhaustive search for the optimal disentangler. This is non-trivial for arbitrary $d$. The approach we take is to generate every two-qudit Clifford tableau, and then transform them into a canonical form by applying single-qubit gates systematically. After reduction to a canonical form, duplicate gates with identical entanglement structure may be removed, leaving the unique disentanglers. Although this procedure is sufficient, the size of the set of stabilizer tableaus grows exponentially with $d$, making $d>3$ beyond the scope of this work. Development of a more efficient search process for larger $d$ remains an open topic of investigation for future work.

\section{Results} \label{sec:results}
\begin{figure}
    \centering
    \includegraphics{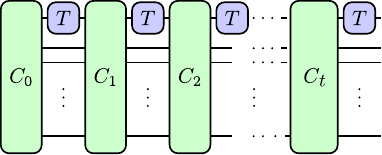}
    \caption{The $T$-doped random Clifford circuits simulated. Each layer consists of a random Clifford operation, followed by a $T$ gate on the first qubit. This is repeated $t$ times, with the circuit depth being the number of layers, which is equal to $t$, the number of non-Clifford $T$ gates in the circuit.}
    \label{fig:random_circ}
\end{figure}

To benchmark the simulation efficiency and scalability of GCAMPS, we performed simulations of random $T$-doped Clifford circuits. A schematic diagram of these circuits is shown in Figure~\ref{fig:random_circ}. Each circuit consists of a randomly chosen Clifford operation, followed by a single $T$ gate. We note that such circuits demonstrate interesting behaviour~\cite{magni_anticoncentration_2025,aditya_mpemba_2025}, and are challenging to classically simulate using conventional techniques. For example, in a state vector simulator, the memory and runtime scale exponentially with the number of qubits. In the case of the  extended stabilizer simulation methods, memory and runtime scale exponentially with the number of $T$ gates, restricting to only low depth circuits. On the other hand, tensor network based methods scale exponentially with the amount of entanglement present in the system. The Clifford group contains the entangling CNOT gate --- and so each randomly chosen Clifford is likely to be highly entangling. Therefore, the bond dimension required to represent the state produced by this circuit scales exponentially with increasing system size. Although, this benchmarking circuit has been extensively studied in qubit systems~\cite{nakhl_stabilizer_2025, fux_disentangling_2025, liu_classical_2025}, there is no study to-date which evaluates the performance for qudit systems.

This prohibitive cost with conventional simulators makes $T$-doped random Cliffords a natural choice to benchmark the performance of a qudit stabilizer tensor network simulator. Previous work has shown that for qubits ($d=2$), hybrid stabilizer tensor network simulation offers a significant speedup for this class of circuit~\cite{nakhl_stabilizer_2025,fux_disentangling_2025}.

\begin{figure*}
    \centering
    \includegraphics[width=1\linewidth]{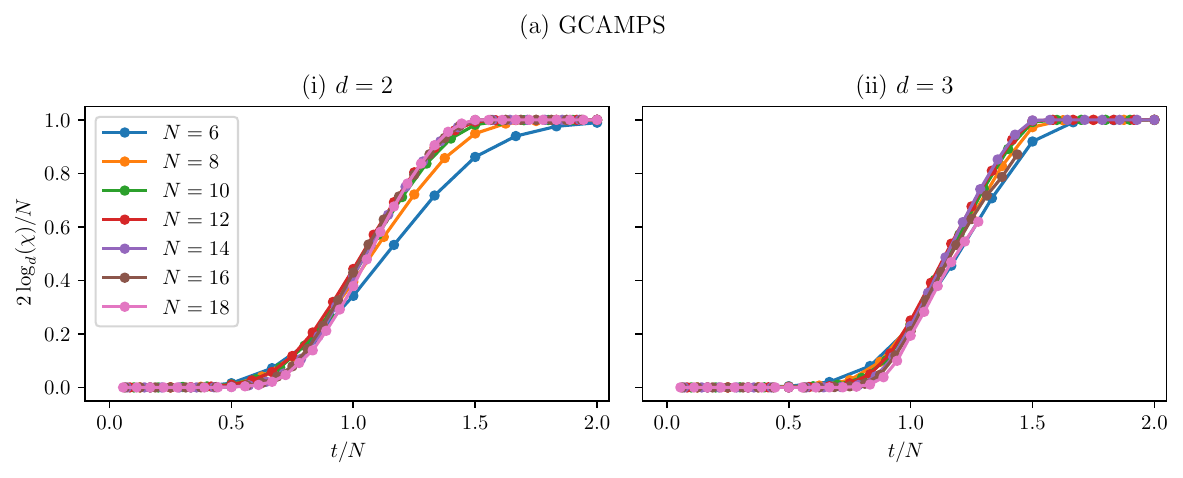}
    \includegraphics[width=1\linewidth]{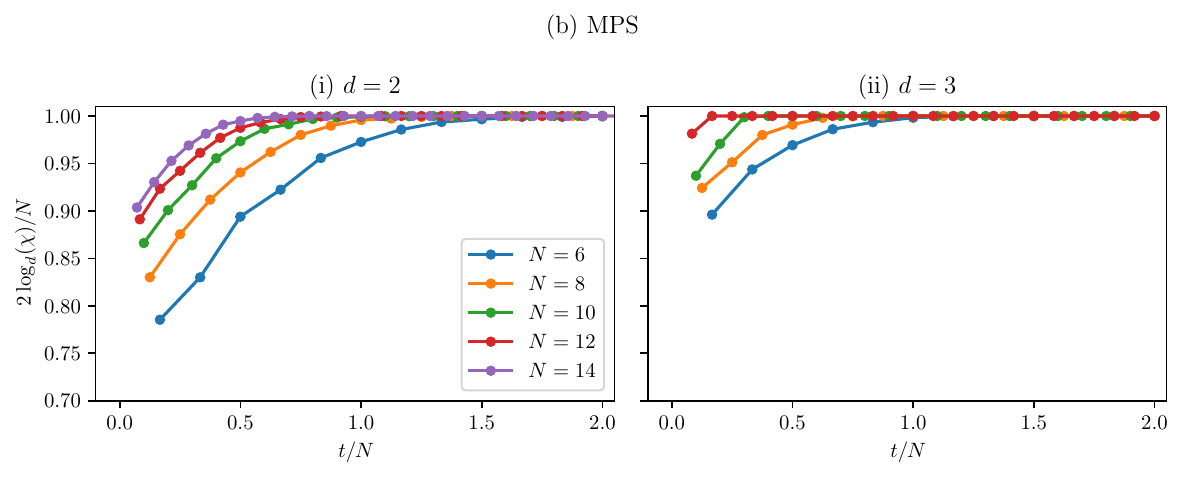}
    \caption{(a) The scaled bond-dimension after each layer for of both (i) qubit and (ii) qutrit $T$-doped random Clifford circuits simulated with GCAMPS as a function of the scaled number of Clifford + T layers $t$. The bond dimension can be though of in analogy with the memory cost associated with the MPS as discussed in Section~\ref{sec:mem}. We note a similar scaling behaviour for both qubits and qutrits, with the qubit simulation exhibiting a phase transition at a lower circuit depth as discussed in Section~\ref{sec:results}. Note that due to computational constraints the $N=16, 18$ qutrit simulations were only simulated up to $23$ layers. (b) Conventional MPS simulation for up to $N=12$, we note that the bond dimension of the MPS saturates almost immediately after very few layers.}
    \label{fig:results}
\end{figure*}

In Figure~\ref{fig:results} we present the results of the $T$-doped random Clifford circuits for both qubit and qutrit GCAMPS and conventional MPS. These results show that, based on the investigated random circuits, we achieve a bond dimension scaling similar to qubit for qutrit systems with GCAMPS, with both cases having a constant bond dimension regime and hence a linear scaling of memory usage with respect to system size. Interestingly, there appears to be a distinction between the qubit and qutrit systems with regard to where the exponential growth regime begins, this appears to be a function of the criterion of when a disentangler is applied in (G)CAMPS with a bond-dimension criterion as used in this work exhibiting this slightly less than optimal scaling compared to other work investigating the same benchmarking circuit with Clifford augmented methods~\cite{nakhl_stabilizer_2025,fux_disentangling_2025}. In comparison, the MPS simulations for both qubits and qutrits in Figure~\ref{fig:results}(b) saturate at the maximal bond dimension after the application of very few layers, consistent with the expectation one may have regarding the simulation of these deep highly entangled circuits.

\subsection{Simulation Cost}
The classical resources required for simulation are directly related to the bond dimension of the tensor network. Hence we expect that GCAMPS, with its lower bond dimension, is significantly more efficient to simulate.

\subsubsection{Execution Time}
\begin{figure*}
    \centering
    \includegraphics[width=\linewidth]{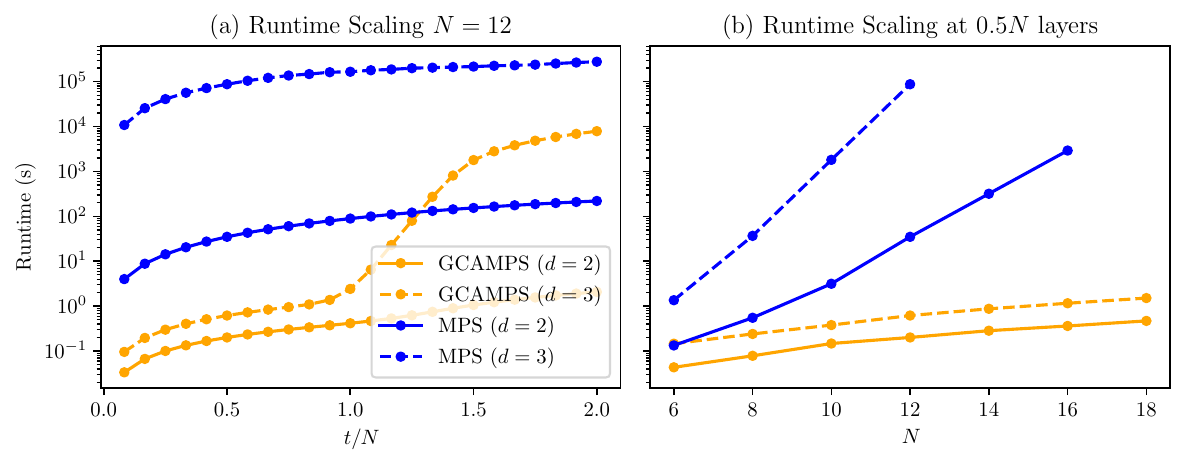}
    \caption{The runtime associated with the execution of the random Clifford + T circuits, (a) for a fixed number of sites ($N=12$), and (b) for a fixed number of layers ($0.5N$). Note that the GCAMPS simulations continue to significantly outperform MPS even at saturated bond dimension as the circuits benchmarked contain many two-qudit operations whereas in GCAMPS the number of two qudit operations on the MPS is at most $N$ for each T gate in addition to those associated with the disentangling procedure which is likely to terminate early for highly non-Clifford systems.}
    \label{fig:sim_time}
\end{figure*}

In Figure~\ref{fig:sim_time}(a) we show the average simulation time per shot as a function of circuit depth for a system of $12$ qudits. This system size was chosen as it is the limit of what can be achieved with conventional MPS simulation within a reasonable timeframe. These results show that there is a significant speedup for GCAMPS simulation in the regime studied compared to a conventional MPS simulation. We find a similar phase transition to that shown in Figure~\ref{fig:results} at about $N$ layers which is a result of the exponential increase in bond-dimension requiring the decomposition of exponentially larger matrices. As the qubit matrices which require decomposition are of size at most $128\times 128$ compared to the qutrit $2187\times2187$ this effect is far less pronounced in the former case. In the low $T$ count regime, as shown in Figure~\ref{fig:sim_time}(b), we find that the improvement in runtime from a conventional MPS simulation is greater for qutrits than for qubits as result of GCAMPS. This is a result of the MPS in GCAMPS being close to a product state in this regime for both qubits and qutrits, making the matrices to be decomposed after each operation of similar size. However, as noted above, for the conventional MPS simulation the qutrit matrices to be decomposed are much larger than those found in the qubit case. 

Note that the simulations shots were run on a single core, however, the main simulation cost is in linear algebra operations, which when using the OpenBLAS backend can be somewhat parallelised for large matrices in particular. Some of the underlying operations in the MPS may be further parallelised or offloaded to a GPU.

\subsubsection{Memory Usage} \label{sec:mem}
\begin{figure*}
    \centering
    \includegraphics[width=1\linewidth]{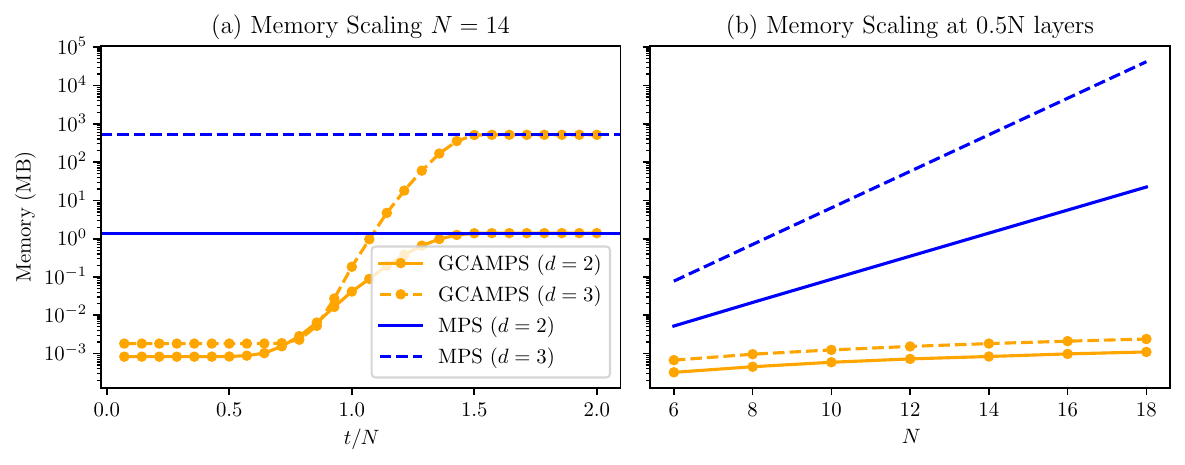}
    \caption{The memory required to simulate the random circuits for both GCAMPS and conventional MPS (a) per layer and (b) per system size. Unlike Figure~\ref{fig:results} which only considers the bond-dimension of the system at the end of each layer here we take into account the fact that the bond-dimension and hence memory will increase during the execution of each circuit layer. As a result, we may reasonably conclude that the conventional MPS will always reach the maximal bond dimension at some point during the execution of the very first layer, and likewise in each layer thereafter. For GCAMPS, we take into account the effect of the non-Clifford operation before the optimisation routine has taken place by noting that the worst case scenario is that the bond dimension between all sites increases by a factor of $d$.}
    \label{fig:mem}
\end{figure*}

In Figure~\ref{fig:mem} we extrapolate from the bond-dimension to get the memory usage required for GCAMPS and conventional MPS. We note that each tensor in the MPS contain $\chi_{l} \times \chi_r \times d$ complex numbers where $\chi_l$ and $\chi_r$ are the bond dimensions to the left and right of the tensor respectively. Using the fact that the complex numbers are represented using two $64$ bit floating point numbers we can easily extrapolate out the memory cost. Unlike in Figure~\ref{fig:results} we aim to account for increases in memory usage throughout the execution of the simulation. Hence for GCAMPS we take the worst case scenario of each $T$ gate increasing the bond dimension at every edge by $d$. For MPS we simply note that the simulation will saturate the bond dimension at some point from the very first layer resulting in a flat memory cost for a system with a fixed number of sites. 

If we note the memory usage for low depth circuits, e.g. at $0.5N$ layers as per Figure~\ref{fig:mem}(b) we find a significant advantage with GCAMPS compared to conventional MPS simulation. This is in line with the bond-dimension results from earlier, and as is the case for runtime, we find a greater improvement in the qutrit memory cost compared to the qubit case with respect their respective conventional MPS simulations. This is attributable to the higher maximal bond dimension (and hence memory) seen with MPS when it comes to qutrits $\chi_{\text{max}}=3^{N/2}$ compared to qubits $\chi_{\text{max}}=2^{N/2}$, whereas in this regime with GCAMPS the typical bond dimensions are $\approx3$ for qutrits and $\approx2$ for qubits which are far closer. 

\section{Conclusion}
In this paper, we present scalable hybrid quantum simulation methods extended to the simulation of qudits --- quantum systems with $d$ levels. Although most quantum computing research is concerned with qubits ($d=2$), many quantum systems that one may be interested in simulating are not necessarily spin-$\frac{1}{2}$, hence extending classical simulation techniques to qudits continues to be an active area of research~\cite{li_complete_2025}. Our results show that for a standard benchmarking circuit on qutrits, Generalised Clifford Augmented Matrix Product State (GCAMPS) simulation offers a significant speed up over traditional classical simulation methods, allowing simulation of circuit depths previously classically inaccessible. Indeed, due to the increased computational complexity of qutrit simulation over qubit simulation, we find that the memory and runtime improvements with GCAMPS for qutrits surpass that of qubits on the same class of benchmarking circuit. By generalising previously developed classical simulation techniques for qubits to qudits, our work allows easy and efficient classical simulation of $d$ level quantum system. In particular, one may use this simulation method in the study of spin-one systems such as the bilinear-biquadratic spin chain~\cite{lauchli_spin_2006} and investigate topological phase transitions beyond spin-1/2 systems. 

\bibliographystyle{ACM-Reference-Format}
\bibliography{references,references2,references3,referencesFree} 

@article{masot-llima_stabilizer_2024,
	title = {Stabilizer {Tensor} {Networks}: {Universal} {Quantum} {Simulator} on a {Basis} of {Stabilizer} {States}},
	volume = {133},
	issn = {0031-9007, 1079-7114},
	shorttitle = {Stabilizer {Tensor} {Networks}},
	url = {https://link.aps.org/doi/10.1103/PhysRevLett.133.230601},
	doi = {10.1103/PhysRevLett.133.230601},
	language = {en},
	number = {23},
	urldate = {2025-10-20},
	journal = {Physical Review Letters},
	author = {Masot-Llima, Sergi and Garcia-Saez, Artur},
	month = dec,
	year = {2024},
	pages = {230601},
}

@article{qian_augmenting_2024,
	title = {Augmenting {Density} {Matrix} {Renormalization} {Group} with {Clifford} {Circuits}},
	volume = {133},
	issn = {0031-9007, 1079-7114},
	url = {https://link.aps.org/doi/10.1103/PhysRevLett.133.190402},
	doi = {10.1103/PhysRevLett.133.190402},
	language = {en},
	number = {19},
	urldate = {2025-10-20},
	journal = {Physical Review Letters},
	author = {Qian, Xiangjian and Huang, Jiale and Qin, Mingpu},
	month = nov,
	year = {2024},
	pages = {190402},
}

@article{pashayan_fast_2022,
	title = {Fast {Estimation} of {Outcome} {Probabilities} for {Quantum} {Circuits}},
	volume = {3},
	issn = {2691-3399},
	url = {https://link.aps.org/doi/10.1103/PRXQuantum.3.020361},
	doi = {10.1103/PRXQuantum.3.020361},
	language = {en},
	number = {2},
	urldate = {2025-09-20},
	journal = {PRX Quantum},
	author = {Pashayan, Hakop and Reardon-Smith, Oliver and Korzekwa, Kamil and Bartlett, Stephen D.},
	month = jun,
	year = {2022},
	pages = {020361},
}

@article{mortier_tensor_2022,
	title = {Tensor {Networks} {Can} {Resolve} {Fermi} {Surfaces}},
	volume = {129},
	issn = {0031-9007, 1079-7114},
	url = {https://link.aps.org/doi/10.1103/PhysRevLett.129.206401},
	doi = {10.1103/PhysRevLett.129.206401},
	language = {en},
	number = {20},
	urldate = {2025-09-20},
	journal = {Physical Review Letters},
	author = {Mortier, Quinten and Schuch, Norbert and Verstraete, Frank and Haegeman, Jutho},
	month = nov,
	year = {2022},
	pages = {206401},
}

@book{ran_tensor_2020,
	address = {Cham},
	series = {Lecture {Notes} in {Physics}},
	title = {Tensor {Network} {Contractions}: {Methods} and {Applications} to {Quantum} {Many}-{Body} {Systems}},
	volume = {964},
	copyright = {https://creativecommons.org/licenses/by/4.0},
	isbn = {978-3-030-34488-7 978-3-030-34489-4},
	shorttitle = {Tensor {Network} {Contractions}},
	url = {http://link.springer.com/10.1007/978-3-030-34489-4},
	language = {en},
	urldate = {2025-09-20},
	publisher = {Springer International Publishing},
	author = {Ran, Shi-Ju and Tirrito, Emanuele and Peng, Cheng and Chen, Xi and Tagliacozzo, Luca and Su, Gang and Lewenstein, Maciej},
	year = {2020},
	doi = {10.1007/978-3-030-34489-4},
}

@misc{liu_classical_2025,
	title = {Classical simulability of {Clifford}+{T} circuits with {Clifford}-augmented matrix product states},
	url = {http://arxiv.org/abs/2412.17209},
	doi = {10.48550/arXiv.2412.17209},
	urldate = {2025-08-27},
	publisher = {arXiv},
	author = {Liu, Zejun and Clark, Bryan K.},
	month = aug,
	year = {2025},
	note = {arXiv:2412.17209 [quant-ph]},
}

@misc{goh_lie-algebraic_2025,
	title = {Lie-algebraic classical simulations for quantum computing},
	url = {http://arxiv.org/abs/2308.01432},
	doi = {10.48550/arXiv.2308.01432},
	urldate = {2025-08-27},
	publisher = {arXiv},
	author = {Goh, Matthew L. and Larocca, Martin and Cincio, Lukasz and Cerezo, M. and Sauvage, Frédéric},
	month = mar,
	year = {2025},
	note = {arXiv:2308.01432 [quant-ph]},
}

@misc{gottesman_stabilizer_1997,
	title = {Stabilizer {Codes} and {Quantum} {Error} {Correction}},
	url = {http://arxiv.org/abs/quant-ph/9705052},
	doi = {10.48550/arXiv.quant-ph/9705052},
	urldate = {2025-08-27},
	publisher = {arXiv},
	author = {Gottesman, Daniel},
	month = may,
	year = {1997},
	note = {arXiv:quant-ph/9705052},
}

@article{bravyi_simulation_2019,
	title = {Simulation of quantum circuits by low-rank stabilizer decompositions},
	volume = {3},
	issn = {2521-327X},
	url = {http://arxiv.org/abs/1808.00128},
	doi = {10.22331/q-2019-09-02-181},
	urldate = {2025-08-27},
	journal = {Quantum},
	author = {Bravyi, Sergey and Browne, Dan and Calpin, Padraic and Campbell, Earl and Gosset, David and Howard, Mark},
	month = sep,
	year = {2019},
	note = {arXiv:1808.00128 [quant-ph]},
	pages = {181},
}

@article{bravyi_improved_2016,
	title = {Improved classical simulation of quantum circuits dominated by {Clifford} gates},
	volume = {116},
	issn = {0031-9007, 1079-7114},
	url = {http://arxiv.org/abs/1601.07601},
	doi = {10.1103/PhysRevLett.116.250501},
	number = {25},
	urldate = {2025-08-27},
	journal = {Physical Review Letters},
	author = {Bravyi, Sergey and Gosset, David},
	month = jun,
	year = {2016},
	note = {arXiv:1601.07601 [quant-ph]},
	pages = {250501},
}

@misc{xu_herculean_2023,
	title = {A {Herculean} task: {Classical} simulation of quantum computers},
	shorttitle = {A {Herculean} task},
	url = {http://arxiv.org/abs/2302.08880},
	doi = {10.48550/arXiv.2302.08880},
	urldate = {2025-08-27},
	publisher = {arXiv},
	author = {Xu, Xiaosi and Benjamin, Simon and Sun, Jinzhao and Yuan, Xiao and Zhang, Pan},
	month = feb,
	year = {2023},
	note = {arXiv:2302.08880 [quant-ph]},
}

@article{turkeshi_magic_2025,
	title = {Magic spreading in random quantum circuits},
	volume = {16},
	issn = {2041-1723},
	url = {http://arxiv.org/abs/2407.03929},
	doi = {10.1038/s41467-025-57704-x},
	number = {1},
	urldate = {2025-07-04},
	journal = {Nature Communications},
	author = {Turkeshi, Xhek and Tirrito, Emanuele and Sierant, Piotr},
	month = mar,
	year = {2025},
	note = {arXiv:2407.03929 [quant-ph]},
	pages = {2575},
}

@misc{feng_quon_2025,
	title = {Quon {Classical} {Simulation}: {Unifying} {Clifford}, {Matchgates} and {Entanglement}},
	shorttitle = {Quon {Classical} {Simulation}},
	url = {http://arxiv.org/abs/2505.07804},
	doi = {10.48550/arXiv.2505.07804},
	urldate = {2025-07-04},
	publisher = {arXiv},
	author = {Feng, Zixuan and Liu, Zhengwei and Lu, Fan and Wang, Ningfeng},
	month = may,
	year = {2025},
	note = {arXiv:2505.07804 [quant-ph]},
}

@misc{frau_stabilizer_2024,
	title = {Stabilizer disentangling of conformal field theories},
	url = {http://arxiv.org/abs/2411.11720},
	doi = {10.48550/arXiv.2411.11720},
	urldate = {2025-07-04},
	publisher = {arXiv},
	author = {Frau, Martina and Tarabunga, Poetri Sonya and Collura, Mario and Tirrito, Emanuele and Dalmonte, Marcello},
	month = nov,
	year = {2024},
	note = {arXiv:2411.11720 [quant-ph]},
}

@article{fan_disentangling_2025,
	title = {Disentangling critical quantum spin chains with {Clifford} circuits},
	volume = {111},
	issn = {2469-9950, 2469-9969},
	url = {http://arxiv.org/abs/2411.12683},
	doi = {10.1103/PhysRevB.111.085121},
	number = {8},
	urldate = {2025-07-04},
	journal = {Physical Review B},
	author = {Fan, Chaohui and Qian, Xiangjian and Zhang, Hua-Chen and Huang, Rui-Zhen and Qin, Mingpu and Xiang, Tao},
	month = feb,
	year = {2025},
	note = {arXiv:2411.12683 [quant-ph]},
	pages = {085121},
}

@misc{zhang_quantum_2024,
	title = {Quantum magic dynamics in random circuits},
	url = {http://arxiv.org/abs/2410.21128},
	doi = {10.48550/arXiv.2410.21128},
	urldate = {2025-03-20},
	publisher = {arXiv},
	author = {Zhang, Yuzhen and Gu, Yingfei},
	month = oct,
	year = {2024},
	note = {arXiv:2410.21128 [quant-ph]},
}

@article{nakhl_calibrating_2024,
	title = {Calibrating the role of entanglement in variational quantum circuits},
	volume = {109},
	issn = {2469-9926, 2469-9934},
	url = {https://link.aps.org/doi/10.1103/PhysRevA.109.032413},
	doi = {10.1103/PhysRevA.109.032413},
	language = {en},
	number = {3},
	urldate = {2025-01-05},
	journal = {Physical Review A},
	author = {Nakhl, Azar C. and Quella, Thomas and Usman, Muhammad},
	month = mar,
	year = {2024},
	pages = {032413},
}

@article{orus_practical_2014,
	title = {A practical introduction to tensor networks: {Matrix} product states and projected entangled pair states},
	volume = {349},
	issn = {00034916},
	shorttitle = {A practical introduction to tensor networks},
	url = {https://linkinghub.elsevier.com/retrieve/pii/S0003491614001596},
	doi = {10.1016/j.aop.2014.06.013},
	language = {en},
	urldate = {2023-08-30},
	journal = {Annals of Physics},
	author = {Orús, Román},
	month = oct,
	year = {2014},
	pages = {117--158},
}

@article{schollwock_density-matrix_2011,
	title = {The density-matrix renormalization group in the age of matrix product states},
	volume = {326},
	issn = {00034916},
	url = {https://linkinghub.elsevier.com/retrieve/pii/S0003491610001752},
	doi = {10.1016/j.aop.2010.09.012},
	language = {en},
	number = {1},
	urldate = {2023-08-30},
	journal = {Annals of Physics},
	author = {Schollwöck, Ulrich},
	month = jan,
	year = {2011},
	pages = {96--192},
}

@book{nielsen_quantum_2012,
	edition = {1},
	title = {Quantum {Computation} and {Quantum} {Information}: 10th {Anniversary} {Edition}},
	isbn = {978-1-107-00217-3 978-0-511-97666-7},
	shorttitle = {Quantum {Computation} and {Quantum} {Information}},
	url = {https://www.cambridge.org/core/product/identifier/9780511976667/type/book},
	urldate = {2023-08-30},
	publisher = {Cambridge University Press},
	author = {Nielsen, Michael A. and Chuang, Isaac L.},
	month = jun,
	year = {2012},
	doi = {10.1017/CBO9780511976667},
}

@article{orus_tensor_2019,
	title = {Tensor networks for complex quantum systems},
	volume = {1},
	issn = {2522-5820},
	url = {http://arxiv.org/abs/1812.04011},
	doi = {10.1038/s42254-019-0086-7},
	number = {9},
	urldate = {2023-08-03},
	journal = {Nature Reviews Physics},
	author = {Orus, Roman},
	month = aug,
	year = {2019},
	note = {arXiv:1812.04011 [cond-mat, physics:hep-lat, physics:quant-ph]},
	pages = {538--550},
}

@phdthesis{dang_distributed_2017,
	title = {Distributed {Matrix} {Product} {State} {Simulations} of {Large}-{Scale} {Quantum} {Circuits}},
	school = {The University of Melbourne},
	author = {Dang, Aidan},
	year = {2017},
}

@article{li_complete_2025,
	title = {A {Complete} and {Natural} {Rule} {Set} for {Multi}-{Qutrit} {Clifford} {Circuits}},
	volume = {426},
	issn = {2075-2180},
	url = {http://arxiv.org/abs/2508.14670},
	doi = {10.4204/EPTCS.426.2},
	urldate = {2025-09-20},
	journal = {Electronic Proceedings in Theoretical Computer Science},
	author = {Li, Sarah Meng and Mosca, Michele and Ross, Neil J. and Wetering, John van de and Zhao, Yuming},
	month = aug,
	year = {2025},
	note = {arXiv:2508.14670 [cs]},
	pages = {23--78},
}

@misc{magni_quantum_2025,
	title = {Quantum {Complexity} and {Chaos} in {Many}-{Qudit} {Doped} {Clifford} {Circuits}},
	url = {http://arxiv.org/abs/2506.02127},
	doi = {10.48550/arXiv.2506.02127},
	urldate = {2025-09-20},
	publisher = {arXiv},
	author = {Magni, Beatrice and Turkeshi, Xhek},
	month = jul,
	year = {2025},
	note = {arXiv:2506.02127 [quant-ph]},
}

@article{wang_stabilizer_2023,
	title = {Stabilizer {Rényi} entropy on qudits},
	volume = {22},
	issn = {1573-1332},
	url = {https://link.springer.com/10.1007/s11128-023-04186-9},
	doi = {10.1007/s11128-023-04186-9},
	language = {en},
	number = {12},
	urldate = {2025-09-20},
	journal = {Quantum Information Processing},
	author = {Wang, Yiran and Li, Yongming},
	month = dec,
	year = {2023},
	pages = {444},
}

@article{lauchli_spin_2006,
	title = {Spin nematics correlations in bilinear-biquadratic {S} = 1 spin chains},
	volume = {74},
	copyright = {http://link.aps.org/licenses/aps-default-license},
	issn = {1098-0121, 1550-235X},
	url = {https://link.aps.org/doi/10.1103/PhysRevB.74.144426},
	doi = {10.1103/PhysRevB.74.144426},
	language = {en},
	number = {14},
	urldate = {2025-09-18},
	journal = {Physical Review B},
	author = {Läuchli, Andreas and Schmid, Guido and Trebst, Simon},
	month = oct,
	year = {2006},
	pages = {144426},
}

@article{qian_clifford_2025,
	title = {Clifford {Circuits} {Augmented} {Time}-{Dependent} {Variational} {Principle}},
	volume = {134},
	copyright = {https://link.aps.org/licenses/aps-default-license},
	issn = {0031-9007, 1079-7114},
	url = {https://link.aps.org/doi/10.1103/PhysRevLett.134.150404},
	doi = {10.1103/physrevlett.134.150404},
	language = {en},
	number = {15},
	urldate = {2025-07-23},
	journal = {Physical Review Letters},
	author = {Qian, Xiangjian and Huang, Jiale and Qin, Mingpu},
	month = apr,
	year = {2025},
	note = {Publisher: American Physical Society (APS)},
}

@article{mello_clifford_2025,
	title = {Clifford {Dressed} {Time}-{Dependent} {Variational} {Principle}},
	volume = {134},
	copyright = {https://link.aps.org/licenses/aps-default-license},
	issn = {0031-9007, 1079-7114},
	url = {https://link.aps.org/doi/10.1103/PhysRevLett.134.150403},
	doi = {10.1103/physrevlett.134.150403},
	language = {en},
	number = {15},
	urldate = {2025-07-23},
	journal = {Physical Review Letters},
	author = {Mello, Antonio Francesco and Santini, Alessandro and Lami, Guglielmo and De Nardis, Jacopo and Collura, Mario},
	month = apr,
	year = {2025},
	note = {Publisher: American Physical Society (APS)},
}

@misc{fux_disentangling_2025,
	title = {Disentangling unitary dynamics with classically simulable quantum circuits},
	url = {http://arxiv.org/abs/2410.09001},
	doi = {10.48550/arXiv.2410.09001},
	urldate = {2025-07-23},
	publisher = {arXiv},
	author = {Fux, Gerald E. and Béri, Benjamin and Fazio, Rosario and Tirrito, Emanuele},
	month = may,
	year = {2025},
	note = {arXiv:2410.09001 [quant-ph]},
}

@article{nakhl_stabilizer_2025,
	title = {Stabilizer {Tensor} {Networks} with {Magic} {State} {Injection}},
	volume = {134},
	copyright = {https://link.aps.org/licenses/aps-default-license},
	issn = {0031-9007, 1079-7114},
	url = {https://link.aps.org/doi/10.1103/PhysRevLett.134.190602},
	doi = {10.1103/physrevlett.134.190602},
	language = {en},
	number = {19},
	urldate = {2025-07-23},
	journal = {Physical Review Letters},
	author = {Nakhl, Azar C. and Harper, Ben and West, Maxwell and Dowling, Neil and Sevior, Martin and Quella, Thomas and Usman, Muhammad},
	month = may,
	year = {2025},
	note = {Publisher: American Physical Society (APS)},
}

@misc{lami_quantum_2024,
	title = {Quantum {State} {Designs} with {Clifford} {Enhanced} {Matrix} {Product} {States}},
	url = {http://arxiv.org/abs/2404.18751},
	urldate = {2024-10-31},
	publisher = {arXiv},
	author = {Lami, Guglielmo and Haug, Tobias and Nardis, Jacopo De},
	month = oct,
	year = {2024},
	note = {arXiv:2404.18751 [quant-ph]},
}

@article{werner_positive_2016_2,
	title = {Positive Tensor Network Approach for Simulating Open Quantum Many-Body Systems},
	volume = {116},
	issn = {0031-9007, 1079-7114},
	url = {http://arxiv.org/abs/1412.5746},
	doi = {10.1103/PhysRevLett.116.237201},
	number = {23},
	urldate = {2023-08-03},
	journal = {Physical Review Letters},
	author = {Werner, A. H. and Jaschke, D. and Silvi, P. and Kliesch, M. and Calarco, T. and Eisert, J. and Montangero, S.},
	month = jun,
	year = {2016},
	note = {arXiv:1412.5746 [cond-mat, physics:quant-ph]},
	pages = {237201},
}

@article{aaronson_improved_2004_2,
	title = {Improved simulation of stabilizer circuits},
	volume = {70},
	issn = {1050-2947, 1094-1622},
	url = {http://arxiv.org/abs/quant-ph/0406196},
	doi = {10.1103/PhysRevA.70.052328},
	number = {5},
	urldate = {2021-10-15},
	journal = {Physical Review A},
	author = {Aaronson, Scott and Gottesman, Daniel},
	month = nov,
	year = {2004},
	note = {arXiv: quant-ph/0406196},
	pages = {052328},
}

@book{nebe2006self,
  title={Self-dual codes and invariant theory},
  author={Nebe, Gabriele and Rains, Eric M and Sloane, Neil JA},
  volume={17},
  year={2006},
  publisher={Springer Science \& Business Media}
}

@book{brandl_efficient_2024,
	title = {Efficient and {Noise}-aware {Stabilizer} {Tableau} {Simulation} of {Qudit} {Clifford} {Circuits} / submitted by {Nina} {Brandl}},
	url = {http://epub.jku.at/obvulihs/10276902},
	language = {en},
	urldate = {2025-09-15},
	author = {Brandl, Nina},
	year = {2024},
}

@software{Nguyen_Sdim_A_Qudit_2025,
author = {Nguyen, Steven and Kabir, Adeeb},
month = feb,
title = {{Sdim: A Qudit Stabilizer Simulator}},
url = {https://github.com/events555/sdim},
version = {1.3.0},
year = {2025}
}

@article{Wang_2020,
   title={Qudits and High-Dimensional Quantum Computing},
   volume={8},
   ISSN={2296-424X},
   url={http://dx.doi.org/10.3389/fphy.2020.589504},
   DOI={10.3389/fphy.2020.589504},
   journal={Frontiers in Physics},
   publisher={Frontiers Media SA},
   author={Wang, Yuchen and Hu, Zixuan and Sanders, Barry C. and Kais, Sabre},
   year={2020},
   month=nov }

@article{magni_anticoncentration_2025,
	title = {Anticoncentration in {Clifford} {Circuits} and {Beyond}: {From} {Random} {Tensor} {Networks} to {Pseudomagic} {States}},
	volume = {15},
	shorttitle = {Anticoncentration in {Clifford} {Circuits} and {Beyond}},
	url = {https://link.aps.org/doi/10.1103/p8dn-glcw},
	doi = {10.1103/p8dn-glcw},
	number = {3},
	urldate = {2025-12-15},
	journal = {Physical Review X},
	author = {Magni, Beatrice and Christopoulos, Alexios and De Luca, Andrea and Turkeshi, Xhek},
	month = sep,
	year = {2025},
	note = {Publisher: American Physical Society},
	pages = {031071},
}

@misc{aditya_mpemba_2025,
	title = {Mpemba {Effects} in {Quantum} {Complexity}},
	url = {http://arxiv.org/abs/2509.22176},
	doi = {10.48550/arXiv.2509.22176},
	urldate = {2025-12-15},
	publisher = {arXiv},
	author = {Aditya, Sreemayee and Summer, Alessandro and Sierant, Piotr and Turkeshi, Xhek},
	month = oct,
	year = {2025},
	note = {arXiv:2509.22176 [quant-ph]},
}

\end{document}